\documentstyle[12pt]{article}

\def\half{{\textstyle{1\over2}}}

\let\la=\label  
  
 \def\bd{\begin{document}} \def\ed{\end{document}}
\def\ds{\documentstyle} \let\fr=\frac \let\bl=\bigl \let\br=\bigr
\let\Br=\Bigr \let\Bl=\Bigl
\let\bm=\bibitem
\let\na=\nabla
\let\pa=\partial \let\ov=\overline
\newcommand{\be}{\begin{equation}}
\newcommand{\ee}{\end{equation}}
\def\ba{\begin{array}}
\def\ea{\end{array}}
\newcommand{\ho}[1]{$\, ^{#1}$}
\newcommand{\hoch}[1]{$\, ^{#1}$}
\newcommand{\bea}{\begin{eqnarray}}
\newcommand{\eea}{\end{eqnarray}}
\newcommand{\ra}{\rightarrow}
\newcommand{\lra}{\longrightarrow}
\newcommand{\Lra}{\Leftrightarrow}
\newcommand{\ap}{\alpha^\prime}
\newcommand{\bp}{\tilde \beta^\prime}
\newcommand{\tr}{{\rm tr} }
\newcommand{\Tr}{{\rm Tr} }
\newcommand{\NP}{Nucl. Phys. }
\newcommand{\tamphys}{\it
Center for Theoretical Physics, Department of Physics\\
Texas A\&M University, College Station, Texas 77843--4242}

\begin{document}

\rightline{CTP-TAMU-2/98}
\rightline{RU98-2-B}
\rightline{SU-ITP-98/01}
\rightline{hep-th/9801072}

\vspace{24pt}

\begin{center}
{ \large {\bf g=1 for Dirichlet 0-branes}}

\vspace{24pt}

M.~J.~Duff${}^a$\footnote{
Research supported in part by NSF Grant PHY-9722090.},
James T.~Liu${}^b$\footnote{
Research supported in part by
the U.~S.~Department of Energy under grant no.~DOE-91ER40651-TASKB.}
and J.~Rahmfeld${}^c$\footnote{
Research supported by NSF Grant PHY-9219345.}

\vspace{10pt}

${}^a$ {\tamphys}

\bigskip

${}^b$ {\it Department of Physics, The Rockefeller University\\
1230 York Avenue, New York, NY 10021-6399}

\bigskip

${}^c$ {\it Department of Physics, Stanford University\\
Stanford, CA 94305-4060}

\vspace{24pt}

\underline{ABSTRACT}

\end{center}

Dirichlet $0$-branes, considered as extreme Type $IIA$ black holes with spin
carried by fermionic hair, are shown to have the anomalous gyromagnetic ratio
$g=1$, consistent with their interpretation as Kaluza-Klein modes.

\vfill
\leftline{}

\newpage

%%%%%%%%%%%%%%%%%%%%%%%%%%%%%%%%%%%%%%%%%%%%%%%%%%%%%%%%%%%%%%%%%%%%%%%%%

\section{Dirichlet 0-branes}
\la{Intro}

Dirichlet $0$-branes enjoy a multiple personality. They are believed to be (in
historical order):

1) {\bf The massive Kaluza-Klein (KK) modes associated with compactifying
D=11 supergravity on a circle \cite{Huq}}. As such they carry mass
$M_n=|n|/R$ and charge
$Q_n=n$ where $R$ is the $S^1$ radius. Consequently, they saturate the BPS
bound and belong to short massive multiplets of $D=10,N=2$ supersymmetry
belonging to the $(44+128+84)$-dimensional representation of the $SO(9)$
little group.

2) {\bf Extreme black holes of Type IIA
string theory \cite{Horowitz1}}. Since these $D=10$ extreme black
holes preserve one half of the supersymmetry \cite{Dufflu}, they
also belong to the same short supermultiplets as the Kaluza-Klein modes.
Moreover, they also have the same
mass and charge quantum numbers as the KK modes and may tentatively
be identified with them \cite{Townsend1}, in much the same way as KK string
states may be identified with extreme black holes in $D=4$
\cite{Duffrahmfeld1,Duff}. There is a curious difference, however. In $D=4$,
spherically symmetric black holes may represent spin-$0$ members
of a KK supermultiplet. In $D=10$, the $(44+128+84)$ multiplet has no
$SO(9)$ singlet, even though the Horowitz-Strominger black hole solution
\cite{Horowitz1} is spherically symmetric and preserves half the
supersymmetry in the extreme limit \cite{Dufflu}.
We shall return to this puzzle below.

3) {\bf The p=0 special case of surfaces of dimension p on which open
strings can end \cite{Polchinski}}. These strings obey Dirichlet boundary
conditions, hence the name. Their Type $I$ origin also implies half of
the available supersymmetry.

4) {\bf The partons of the Matrix Model \cite{Banks}}. In this
picture, a charge $n$ Dirichlet $0$-brane is a zero energy bound state of $n$
singly charged $0$-branes \cite{Sethi}.

In the absence of a complete formulation of some overarching $M$-theory, each
of the above pictures is necessary and useful, and it is still worthwhile to
perform consistency checks on their presumed equivalence. One such check
concerns the gyromagnetic ratios, similar to the check performed for black
holes and string states in $D=4$ \cite{Duff:1997bs}. In this paper,
we compare the $g$-factors of Dirichlet $0$-branes from the
perspectives of (1) and (2). The black hole calculation relies on generating
the spin from the fermionic zero-modes as in \cite{Aichelburg,Duff:1997bs}
and yields the anomalous value $g=1$, consistent with the KK result
from compactifying on a circle%
\footnote{In this paper we assume for
simplicity a direct product $M^{10}\times S^{1}$ but, as discussed
in \cite{susy}, the $D=11$ origin of Type $IIA$ string theory
\cite{Howe,Townsend1,Wittenvarious} and the Dirichlet
$0$-brane interpretation work equally
well when the compactifying manifold is a $U(1)$ fibration.}.

\section{g=1 from Kaluza-Klein arguments}

Recall that for a Kaluza-Klein compactification from five to four dimensions,
it was shown in \cite{Hosoya:1984} that all massive Kaluza-Klein states share
a common gyromagnetic ratio of $g=1$.  At first sight, this is somewhat of a
surprise, as tree-level unitarity ordinarily demands that the gyromagnetic
ratio takes on the ``natural'' value of $g=2$
\cite{Weinberg,Ferrara:1992b,Jackiw}.
Consequently, the value $g=2$ is required in QED, in the Standard Model and
indeed in the perturbative sector of open string theory \cite{Argyres}.
However this consideration is only important in the energy range
$M_{\rm Pl} > E > M/Q$.  Since Kaluza-Klein states have $M\sim Q$ (which is
also the BPS condition), we see that this range is essentially
empty, and hence there is no conflict with tree-level unitarity,
gauge invariance or any other principle.

Although the gyromagnetic ratio is traditionally defined in $D=4$ as a
proportionality relation between spin and magnetic moment $3$-vectors, it
turns out that there is a natural generalization to arbitrary dimensions.
To see this, we note that for $D>4$ the angular momentum is more properly
represented not as a vector, $\vec J$, but rather as the adjoint
representation of the $SO(D-1)$ rotation group, $J^{ij}$
($i,j=1,2,\ldots,D-1$).  Similarly, both the magnetic field strength,
$F_{ij}$, and the magnetic dipole moment, $\mu^{ij}$, are in the $2$-index
antisymmetric representation as well, so that the magnetic dipole
interaction in any dimension may be written as
\begin{equation}
\Delta E=-{1\over2}\mu^{ij}F_{ij}\ .
\label{eq:dipoleint}
\end{equation}
The gyromagnetic ratio $g$ may then be defined in general by
\begin{equation}
\mu^{ij} = {gQ\over2M}J^{ij}\ ,
\label{eq:gfactor}
\end{equation}
which reduces to the standard expression in $D=4$.

Given this generalization of the gyromagnetic ratio, it is natural to
wonder whether $g=1$ is a universal property of Kaluza-Klein states in any
dimension.  In fact, since the results of \cite{Hosoya:1984} are
mostly dimension independent, they are easily extended to arbitrary
dimension, thus showing that $g=1$ is indeed universally true.  As an
example, we consider the Kaluza-Klein reduction of a massless $d$-form
potential $C_d$ with field strength $K_{d+1} = dC_d$ from $D+1$ to $D$
dimensions.  We use the Kaluza-Klein decomposition
\begin{equation}
G_{MN} = \pmatrix{g_{\mu\nu}+e^{2\varphi}A_\mu A_\nu&e^{2\varphi}A_\mu\cr
e^{2\varphi}A_\nu&e^{2\varphi}}\ ,
\end{equation}
where for simplicity $e^{2\varphi}$, which determines the length-scale of the
compactification, is taken to be a constant.  Furthermore, in order to
examine the magnetic dipole interaction, it is sufficient to work to linear
order in the Kaluza-Klein gauge field $A_\mu$.  In this case,
the ($D+1$)-dimensional equation of motion, $\nabla^MK_{MN_1\cdots N_d}=0$,
reduces to the following two equations in $D$ dimensions:
\begin{eqnarray}
0 &=& D^\lambda K_{\lambda\mu_1\cdots\mu_d}
-(\nabla^\lambda A_\lambda+A_\lambda \nabla^\lambda -e^{-2\varphi}\partial_i)
K_{i\mu_1\cdots\mu_d}+d\,\nabla^\lambda A_{\mu_1}K_{i\lambda\mu_2\cdots\mu_d}
\ ,\nonumber\\
0 &=& D^\lambda K_{i\lambda\mu_1\cdots\mu_{d-1}}
+\half e^{2\varphi}F^{\rho\sigma}K_{\rho\sigma\mu_1\cdots\mu_{d-1}}\ ,
\end{eqnarray}
where $D_\mu = \nabla_\mu - A_\mu\partial_i$ and antisymmetrization of the
$\{\mu_1,\mu_2,\mu_3,\ldots\}$ is always implied.  Note that $i$ denotes the
compact coordinate
and only takes on a single value in this case.  For $z= z+2\pi L$, the $n$-th
Kaluza-Klein state has mass $M_n=\langle e^{-\varphi}\rangle |n|/L=|n|/R$
measured in the $(D+1)$-dimensional metric%
\footnote{Note that for $D=10$ the usual string dilaton $\phi$ is
given by $\phi = 3\varphi/2$.  This explains the unconventional factors in
the expressions for the mass and radius.}
and charge $Q_n=n/L$. Here $R$ is the radius
$R=L\langle e^{\varphi}\rangle$.

For massive states ($n\ne 0$) we may employ the gauge condition
$C_{i\mu_1\cdots\mu_{d-1}} = 0$ to finally arrive at the equation of motion
\begin{equation}
0 = (D^\lambda D_\lambda - M_n^2)C_{\mu_1\cdots\mu_d}
+\half Q_n F_{\rho\sigma}(\Sigma^{\rho\sigma}C)_{\mu_1\cdots\mu_d}
-i{\textstyle{d\over Q_n}} e^{2\varphi}F^{\rho\sigma}\nabla_{\mu_1}\nabla_\rho
C_{\sigma\mu_2\cdots\mu_d}\ ,
\end{equation}
where the second term contains the magnetic dipole interaction and the last
term represents the Thomas precession.  We have used the explicit form of the
angular momentum generators in the $d$-fold antisymmetric representation:
\begin{equation}
(\Sigma^{\mu\nu})_{\{\alpha\}}{}^{\{\beta\}}
= -2id\delta^{[\mu}{}_{[\alpha_1}\eta^{\nu][\beta_1}
\delta_{\alpha_2\cdots\alpha_d]}^{\beta_2\cdots\beta_d]}\ ,
\end{equation}
where all symbols are antisymmetric with weight one.  Combining
(\ref{eq:dipoleint}) and (\ref{eq:gfactor}), we finally obtain the
result that $g=1$ for the massive Kaluza-Klein modes of the $d$-form
potential $C_d$.  Similar arguments may be used to show that $g=1$ holds for
the reduction of arbitrary fields as well.

The general result of $g=1$ for Kaluza-Klein states can also be understood
from the closed string point of view.  While the left- and right-moving
modes on the string world sheet naturally lead to the definition of
separate $g$-factors \cite{Russo:1995,Senblack1}
\begin{equation}
(g_L,g_R) = \left({2\langle S_R\rangle\over\langle S_L+S_R\rangle},
{2\langle S_L\rangle\over\langle S_L+S_R\rangle}\right)
\end{equation}
($\vec S_L$ and $\vec S_R$ are contributions to the total spin from
worldsheet left- and right-movers), it is nevertheless possible to focus on
a single Kaluza-Klein $g_{KK}$.  This has in fact been calculated
in \cite{Russo:1995} with the result
\begin{equation}
g_{KK}=1-{W\over Q}{\langle S_L-S_R\rangle\over\langle S_L+S_R\rangle}\ ,
\end{equation}
where $Q$ and $W$ are Kaluza-Klein and winding charge respectively.
Therefore we see that $g_{KK}=1$ for pure Kaluza-Klein states where $W=0$.
Note that although \cite{Russo:1995,Senblack1} were interested in four
dimensions, the resulting $g$-factors are in fact valid in arbitrary
dimensions based on the definitions (\ref{eq:dipoleint}) and
(\ref{eq:gfactor}).

\section{g=1 from black hole arguments}

We now turn to the properties of the Dirichlet $0$-brane, viewed as a soliton
of ten-dimensional Type IIA supergravity, and proceed to calculate its
gyromagnetic ratio.  In this case, the relevant part of
the supergravity action is given (in the Einstein frame) by
\begin{equation}
{\cal L}={1\over2\kappa^2}\sqrt{-g}[R-{1\over2}(\partial_\mu\phi)^2
-{1\over4}e^{{3\over2}\phi}F_{\mu\nu}^2]\ ,
\label{eq:IIAaction}
\end{equation}
where $F=dA$ is the field strength of the RR gauge boson $A_\mu$.  Ignoring
both $B_{\mu\nu}$ and $C_{\mu\nu\lambda}$, which do not play a role in the
$0$-brane solution, the relevant supersymmetry variations are given by
\cite{Campbell:1984zc,Huq}
\begin{eqnarray}
\delta e_\mu{}^a&=&-i\overline{\epsilon}\gamma^a\psi_\mu\ ,\nonumber\\
\delta\phi&=&-i\sqrt{2}\overline{\epsilon}\Gamma_{11}\lambda\ ,\nonumber\\
\delta A_\mu&=&-ie^{-{3\over4}\phi}\overline{\epsilon}(\Gamma_{11}\psi_\mu
+{3\over2\sqrt{2}}\gamma_\mu\lambda)\ ,
\end{eqnarray}
for the bosons, and
\begin{eqnarray}
\delta\lambda&=&-{1\over2\sqrt{2}}[\gamma^\mu\Gamma_{11}\partial_\mu\phi
+{3\over8}e^{{3\over4}\phi}\gamma^{\mu\nu}F_{\mu\nu}]\epsilon\ ,\nonumber\\
\delta\psi_\mu&=&[\nabla_\mu-{1\over64}e^{{3\over4}\phi}
(\gamma_\mu{}^{\nu\lambda}-14\delta_\mu{}^\nu\gamma^\lambda)
\Gamma_{11}F_{\nu\lambda}]\epsilon\ ,
\end{eqnarray}
for the fermions.

The $0$-brane soliton was originally constructed in \cite{Horowitz1}
as a ten-dimensional black hole solution to the bosonic equations of motion
of (\ref{eq:IIAaction}). In the extreme limit we obtain the field
configuration
\begin{eqnarray}
&&ds^2=-e^{-{7\over6}(\phi-\phi_0)}dt^2+e^{{1\over6}(\phi-\phi_0)}d\vec x^2\ ,
\nonumber\\
&&A_0= e^{-{4\over3}(\phi-\phi_0)-{3\over4}\phi_0}\ ,\nonumber\\
&&e^{{4\over3}(\phi-\phi_0)}=
\left(1+{ke^{-{3\over4}\phi_0}\over r^7}\right)\ .
\label{eq:D0brane}
\end{eqnarray}
The properties of this solution may be determined asymptotically through an
ADM procedure.  To read off the mass $M$, charge $Q$, angular
momentum $J^{ij}$, electric dipole moment $d^{i}$ and magnetic dipole
moment $\mu^{{ij}}$, we compare the $0$-brane solution with the generic
$D$-dimensional configuration:
\begin{eqnarray}
g_{00}&\sim& -\left(1-{2\kappa^2M\over(D-2)\omega_{D-2}r^{D-3}}\right)
\ ,\nonumber\\
g_{ij}&\sim& \left(1+{2\kappa^2M\over(D-2)(D-3)\omega_{D-2}r^{D-3}}\right)
\delta_{ij}\ ,\nonumber\\
g_{0i}&\sim& -\kappa^2J^{ij}{\hat x_j\over \omega_{D-2}r^{D-2}}\ ,
\label{eq:ADM}
\end{eqnarray}
for the metric (where $\omega_n=2\pi^{(n+1)/2}/\Gamma((n+1)/2)$ is the
volume of an $n$-sphere) \cite{Myers:1986un}, and
\begin{eqnarray}
A_0&\sim&{Q\over(D-3)r^{D-3}}+d^i{\hat x_i\over r^{D-2}}\ ,\nonumber\\
A_i&\sim&\hphantom{ {Q\over(D-3)r^{D-3}} }-\mu^{ij}{\hat x_j\over r^{D-2}}\ ,
\label{eq:multipole}
\end{eqnarray}
for the gauge fields.
Thus the $0$-brane has mass $M={7\omega_8\over2\kappa^2}ke^{-{3\over4}\phi_0}$
and electric charge $Q=-7ke^{-{3\over2}\phi_0}$.
The bosonic $0$-brane solution is spherically symmetric and hence appears
to have vanishing angular momentum and vanishing dipole moments. However,
this is not necessarily the case when fermion zero modes are included, as
we see below. Note that in $D=10$, bosonic Kerr-type angular momentum
associated with {\it rotating} black hole solutions will play no role in
determining the $g$-factor of Dirichlet $0$-branes. The role of Kerr
angular momentum in four dimensions will, however, be discussed in
section (\ref{Lower}).

Comparing the mass with the charge, we find that the $0$-brane saturates the
Bogomol'nyi bound, $M\ge|Z|$, where
$Z={\omega_8\over2\kappa^2}Qe^{{3\over4}\phi_0}$ is
the central charge.  As a result, this state preserves exactly half of the
supersymmetries as may also be seen from the supersymmetry variations
\cite{Dufflu}.  In fact, inserting the solution, (\ref{eq:D0brane}),
into the fermionic supersymmetry variations, we find
\begin{eqnarray}
\delta\lambda&=&-{1\over\sqrt{2}}\gamma^i\Gamma_{11}\partial_i\phi
P_+\epsilon\ ,\nonumber\\
\delta\psi_0&=&-{7\over12}\gamma_0{}^i\partial_i\phi P_+\epsilon\ ,\nonumber\\
\delta\psi_i&=&{1\over12}(\gamma_i{}^j-7\delta_i{}^j)\partial_j\phi
P_+\epsilon\ ,
\label{eq:D0susy}
\end{eqnarray}
where $\epsilon=e^{-{7\over24}\phi}\epsilon_0$ with $\epsilon_0$ a constant
spinor.  These supersymmetry variations clearly indicate the splitting of
the 32 (real) component spinor $\epsilon$ into a set of Killing spinors,
$P_+\epsilon_-=0$, and a set of fermion zero mode spinors,
$P_+\epsilon_+=\epsilon_+$, where the half-supersymmetry projection operator
is given by $P_\pm={1\over2}(1\pm\gamma^{\overline{0}}\Gamma_{11})$ and the
overline indicates tangent space indices.

Before focusing our attention on the gyromagnetic ratio calculation, we note
that it is the presence of the fermion zero modes that ensures that the
$0$-brane fills out a complete supermultiplet.  In the absence of any
fermion zero modes ({\it i.e.}~in a hypothetical bosonic truncation of the
supergravity theory), the bosonic solution of (\ref{eq:D0brane}) would
describe a single spinless particle.  However since the $0$-brane soliton
preserves exactly half of the ten-dimensional $N=2A$ supersymmetry, there
are 16 fermion zero modes present, corresponding to the non-trivially
realized supercharges $Q_+$.  Recall that the supersymmetry algebra in the
rest frame is of the form
\begin{equation}
\{Q_+,Q_+\}=M+|Z|\ ,\qquad \{Q_-,Q_-\}=M-|Z|\ ,\qquad
\{Q_+,Q_-\}=0\ ,
\end{equation}
so that in an appropriate basis the 16 $Q_+$ form a $SO(16)$ Clifford
algebra with two 128-dimensional spinor representations.  Making note of the
special embedding of $SO(9)$ in $SO(16)$ where $16\to16$, $128\to128$ and
$128'\to44+84$, this demonstrates that the D0-brane soliton indeed fills
out a complete $(44+128+84)$-dimensional representation of the massive $SO(9)$
little group.

Perhaps somewhat surprisingly, the Clifford vacuum associated
with the bosonic solution (\ref{eq:D0brane}) must necessarily carry some
non-trivial $SO(9)$ spin.  This may be seen by noting that the 16 $Q_+$'s
can be grouped to form 8 complex pairs of creation and annihilation operators
on the Clifford vacuum.  However there is no 8-dimensional representation of
$SO(9)$.  Hence any choice of the Clifford vacuum on which the
creation/annihilation operators act would break $SO(9)$ invariance.  The most
natural choice allowing complex $Q_+$ appears to follow the decomposition
$SO(9) \supset SU(2)\times SU(4)$ where $16\to(2,4)+(2,\overline{4})$.  In
this case, the Clifford vacuum corresponds to a $SU(2)\times SU(4)$ singlet,
$(1,1)$, while the complex creation operators $Q_+^\dagger$ transform as
$(2,4)$.  With up to eight creation operators acting on the Clifford vacuum,
the only $SU(2)\times SU(4)$ singlets appear as $|\Omega\rangle$ and
$(Q_+^\dagger)^8|\Omega\rangle$, where $|\Omega\rangle$ corresponds to the
bosonic solution (\ref{eq:D0brane}).  Since both the 44 and 84 dimensional
representations of $SO(9)$ contain $SU(2)\times SU(4)$ singlets, we find
that the bosonic solution falls in general in some combination of the 44 and
84 dimensional representations, and that it is not possible to determine
where it truly belongs based purely from the superalgebra alone.
Nevertheless, the complete resulting $0$-brane supermultiplet is in precisely
the correct $(44+128+84)$-dimensional representation necessary to correspond
to the massive Kaluza-Klein states of the $11$-dimensional supergraviton.

Properties of the Dirichlet $0$-brane superpartners may be derived by
performing successive supersymmetry transformations to the original bosonic
solution.  In particular, denoting the bosonic solution by $\Phi$, complete
information of its superpartners are encoded in the finite transformation
\cite{Aichelburg}
\begin{equation}
\Phi\longrightarrow e^\delta\Phi=\Phi+\delta\Phi+{1\over2}\delta\delta\Phi
+\cdots\ .
\end{equation}
With 16 zero modes, this series does not terminate until order $\delta^{16}$.
However, since each additional pair of supersymmetries brings in another
power of the momentum, it is sufficient to work only up to second-order
variations when considering properties of the dipole moments.  This
approach was used in \cite{Duff:1997bs} to calculate both electric
and magnetic dipole moments of four-dimensional $N=4$ black holes.

The double supersymmetry variation of the metric yields corrections to the
mass and angular momentum of the superpartners.  Using $\delta\delta
g_{\mu\nu}=-2i\overline{\epsilon}\gamma_{(\mu}\delta\psi_{\nu)}$ and the
gravitino transformations for the $0$-brane background from (\ref{eq:D0susy}),
we find (for zero-mode spinors $\epsilon_+$)
\begin{equation}
\delta\delta g_{00}=0\ ,\qquad\delta\delta g_{ij}=0\ ,
\end{equation}
so in fact there is no mass shift among the different members of the
supermultiplet.  On the other hand, we may read off the angular momentum
from the asymptotics of the mixed components of the metric
\begin{eqnarray}
\delta\delta g_{0i}&=&-{2i\over3}(\overline{\epsilon}_+\gamma_{0i}{}^j
\epsilon_+)\partial_j\phi\nonumber\\
&\sim&{\kappa^2M\over\omega_8}
(i\overline{\epsilon}_+\gamma_{0i}{}^j\epsilon_+){\hat x_j\over r^8}\ .
\end{eqnarray}
Using the ADM definition of (\ref{eq:ADM}), the resulting angular momentum
is
\begin{equation}
J^{ij}=-M (i\overline{\epsilon}_+\gamma_0{}^{ij}\epsilon_+)\ .
\end{equation}
More precisely, this expression may be used to determine the angular
momentum carried by any member of the $0$-brane supermultiplet by choosing
the fermion zero mode spinors $\epsilon_+$ appropriately to correspond to
the given state.

For the RR gauge field $A_\mu$, we find
\begin{eqnarray}
\delta\delta A_0&=&0\ ,\nonumber\\
\delta\delta A_i&=&{2i\over3}e^{-{3\over4}\phi}
(\overline{\epsilon}_+[\gamma_i{}^j+2\delta_i{}^j]\Gamma_{11}\epsilon_+)
\partial_j\phi\nonumber\\
&\sim&{Q\over2} (i\overline{\epsilon}_+\gamma_{0i}{}^j\epsilon_+)
{\hat x_j\over r^8}+\hbox{gauge}\ .
\end{eqnarray}
The vanishing of $\delta\delta A_0$  indicates that no electric dipole
moment is generated for the superpartners, while $\delta\delta A_i$ results
in a magnetic dipole moment
\begin{equation}
\mu^{ij}=-{Q\over2} (i\overline{\epsilon}_+\gamma_0{}^{ij}\epsilon_+)\ .
\end{equation}
Finally, using the definition (\ref{eq:gfactor}) for the $g$-factor,
we see that indeed $g=1$ for all states in the $0$-brane supermultiplet,
in agreement with the expected Kaluza-Klein result.

\section{Lower dimensions and U-duality}
\la{Lower}

While we have shown above that Kaluza-Klein solitons have $g=1$, this
statement is easily generalized to other $0$-brane states as well through
the inclusion of $U$-duality.  In particular, any $0$-brane (in any dimension)
that is $U$-conjugate to a Kaluza-Klein state must necessarily have $g=1$
as long as $U$-duality is valid.  This result is most straightforward in
the maximal supergravities where all gauge bosons are graviphotons so that
$U$-duality indeed relates all $0$-branes to Kaluza-Klein states.

The bosonic $0$-brane solutions \cite{Horowitz1} are easily
constructed in any dimension, and in fact are part of a large family of
general $p$-brane solutions \cite{Lublack,Khuristring}.
To see this, we recall that $p$-branes may
be constructed as solutions to a general bosonic action of a scalar field,
$\varphi$, coupled to a $(p+2)$-form field strength, $F_{p+2}=dA_{p+1}$, in
the presence of gravity.  Correspondence to actual theories is then
obtained by choosing the scalar coupling appropriately, and with
proper identification of $\varphi$ and $F_{p+2}$ with the actual
supergravity fields.  While this approach is valid for the bosonic
solutions, it does not appear possible to treat the fermions in an equally
general manner.  In particular, since we are interested in the supersymmetric
properties of the $0$-branes, each theory must be investigated individually.

To give further evidence for $g=1$, we now examine $0$-branes in
eight-dimensional $N=2$ supergravity.  This theory has $SL(3)\times SL(2)$
$U$-duality, and has six $1$-form gauge fields transforming as $(3,2)$ under
the $U$-duality group.  Ignoring the $2$-form and $3$-form potentials, the
bosonic part of the action is
\begin{equation}
{\cal L}={1\over2\kappa^2}\sqrt{-g}[R-\tr P_\mu P^\mu
-{1\over2}(\partial_\mu\phi)^2-e^{-2\phi}(\partial_\mu b)^2
-{1\over4}e^\phi F_{\mu\nu}^TMF^{\mu\nu}
-{1\over4}e^{-\phi}G_{\mu\nu}^TMG^{\mu\nu}]\ ,
\label{eq:8action}
\end{equation}
where the seven scalars of the $D=8$ $N=2$ theory have been split into
$(b,e^\phi)$ describing $SL(2)/SO(2)$ and $M=V^TV$ with $V_{ai}$ being
a coset representative of $SL(3)/SO(3)$.  For the latter coset, the
scalar kinetic terms, $P_\mu$, and composite $SO(3)$ connection, $Q_\mu$,
are given by the decomposition of the Maurer-Cartan form,
$(\partial_\mu VV^{-1})^{ab} = P_\mu^{(ab)} +Q_\mu^{[ab]}$.
Note that from an $M$-theory
point of view the six gauge fields split into three Kaluza-Klein fields,
$A_\mu^i$, and three fields coming from the reduction of the original
$3$-form potential, $B_{\mu\,i} \sim \epsilon_{ijk} C_{\mu jk}$.  The
field strengths are then given by $F^i=dA^i$ and $G_i = dB_i + 2b F^i$.

The relevant supersymmetries are \cite{Salam:1985ft}
\begin{eqnarray}
\delta\psi_\mu&=&[\nabla_\mu+{1\over4}Q_\mu^{ab}\Gamma^{ab}
+{1\over2}e^{-\phi}\partial_\mu b\Gamma^{\overline{123}}]\epsilon
\nonumber\\
&&+{1\over48}[e^{{1\over2}\phi}F_{\nu\lambda}{}^i\Gamma_i
-{1\over2}e^{-{1\over2}\phi}
\epsilon^{ijk}G_{\nu\lambda\,i}\Gamma_{jk}][\gamma_\mu{}^{\nu\lambda}
-10\delta_\mu{}^\nu\gamma^\lambda]\epsilon\ ,\nonumber\\
\delta\chi^a&=&[{1\over2}P_\mu^{ab}\Gamma^b+{1\over6}\partial_\mu\phi
\Gamma^a-{1\over6}e^{-\phi}\partial_\mu
b\epsilon^{abc}\Gamma^{bc}]\gamma^\mu\epsilon\nonumber\\
&&-{1\over8}[e^{{1\over2}\phi}F_{\mu\nu}{}^a
+{1\over6}e^{-{1\over2}\phi}\epsilon_{bcd}
G_{\mu\nu\,d}(\Gamma_{abc}-4\delta_{ab}\Gamma_c)]\gamma^{\mu\nu}\epsilon
\ ,\nonumber\\
\delta e_\mu{}^a&=&-\overline{\epsilon}\gamma^a\psi_\mu\ ,\nonumber\\
V_{ai}\delta A_\mu^i&=&-e^{-{1\over2}\phi}\overline{\epsilon}
[\Gamma^a\psi_\mu+{1\over6}\gamma_\mu(\Gamma^a\Gamma^b+6\delta^{ab})\chi^b]
\ ,\nonumber\\
(V^{-1})^{ia}\delta B_{\mu\,i}&=&-4bV_{ai}\delta A_\mu^i
-{1\over2}e^{{1\over2}\phi}\epsilon^{abc}\overline{\epsilon}
[\Gamma^{bc}\psi_\mu
-{1\over6}\gamma_\mu\Gamma^b(\Gamma^c\Gamma^d-12\delta^{cd})\chi^d]\ ,
\label{eq:8susy}
\end{eqnarray}
where we have used an $11$-dimensional notation for the Dirac matrices,
$\Gamma^M = \{\gamma^\mu,\Gamma^i\}$.

The $0$-branes may be constructed by
solving the first-order Killing-spinor equations resulting from setting
the fermion variations in (\ref{eq:8susy}) to zero
\cite{Duff:1997bs}.  For a $0$-brane
charged under the Kaluza-Klein fields $A_\mu^i$, we choose a
half-supersymmetry projection $P_\pm={1\over2}(1\pm\gamma^{\overline{0}}
\hat n\cdot\Gamma)$ where $\hat n$ is a unit 3-vector selecting the $U(1)$
component that the 0-brane is charged under.  Setting $\phi_0=0$ for
simplicity, the $0$-brane solution is then given by
\begin{eqnarray}
&&ds^2=-e^{-{5\over3}\phi}dt^2+e^{{1\over3}\phi}d\vec x^2\ ,\nonumber\\
&&Q_i^{ab}=0\ ,\qquad P_i^{ab}=-{1\over3}\partial_i\phi(\delta^{ab}
-3\hat n^a\hat n^b)\ ,\qquad b=0\ ,\nonumber\\
&&E_i^a=\mp{3\over2}\hat n^a\partial_ie^{-{4\over3}\phi}\ ,
\end{eqnarray}
with resulting Killing-spinor equations
\begin{eqnarray}
\delta\psi_0&=&-{5\over6}\gamma_0{}^i\partial_i\phi
P_+\epsilon\ ,\nonumber\\
\delta\psi_i&=&{1\over6}(\gamma_i{}^j-5\delta_i{}^j)\partial_j\phi
P_+\epsilon\ ,\nonumber\\
\delta\chi^a&=&\hat n^a\gamma^i\gamma^{\overline{0}}\partial_i\phi
P_+\epsilon\ .
\end{eqnarray}
In this case, $\epsilon = e^{-{5\over12}\phi}\epsilon_0$ where $\epsilon_0$
is a constant spinor.  Note that the first order Killing-spinor equations
are incomplete in the sense that they do not fully determine the behavior of
$e^{2\phi}$.  Only when the bosonic equations of motion are taken into
account do we find the harmonic function condition, namely
$e^{2\phi}=1+k/r^5$ for a spherically symmetric 0-brane solution.
In terms of $k$, the 0-brane has mass $M={5\omega_6\over2\kappa^2}k$
and a six-dimensional charge
vector ${\cal Q}= [\vec Q,\vec 0]=[-5k\hat n,\vec 0]$, where the first
entry denotes the
three Kaluza-Klein charges and the last one the three $B_{\mu\,i}$ charges.

Following the procedure carried out above for the Dirichlet $0$-brane, we now
examine double supersymmetry variations to determine the properties of
superpartners.  For the metric, we find
\begin{equation}
\delta\delta g_{0i}=-(\overline{\epsilon}_+\gamma_{0i}{}^j\epsilon_+)
\partial_j\phi\ ,
\end{equation}
resulting in a supersymmetry generated spin $J^{ij}=-M(\overline{\epsilon}_+
\gamma_0{}^{ij}\epsilon_+)$.  The gauge field variations are somewhat more
intricate.  We find
\begin{eqnarray}
\delta\delta A_0^a&=&e^{-{1\over2}\phi}
(\overline{\epsilon}_+ \Gamma^a \gamma_0{}^i \epsilon_+)
\partial_i\phi\ ,\nonumber\\
\delta\delta B_{0\,a}&=&{\textstyle{1\over2}}e^{{1\over2}\phi}\epsilon^{abc}
(\overline{\epsilon}_+
\Gamma^{bc}\gamma_0{}^i\epsilon_+)\partial_i\phi\ ,\nonumber\\
\delta\delta A_i^a&=&\hat n^ae^{{1\over3}\phi}
(\overline{\epsilon}_+\gamma_{0i}{}^j \epsilon_+)
\partial_j\phi + \hbox{gauge}\ ,\nonumber\\
\delta\delta B_{i\,a}&=&\epsilon^{abc}\hat n^b e^{{4\over3}\phi}
(\overline{\epsilon}_+\Gamma^c\gamma_{0i}{}^j\epsilon_+)
\partial_j\phi+\hbox{gauge}\ ,
\end{eqnarray}
indicating that superpartners carry both electric and magnetic dipole
moments.  Using the asymptotics of
(\ref{eq:multipole}), we find the complete electric and magnetic dipole
moment vectors of the $0$-brane to be
\begin{eqnarray}
d^i &=& [
{|Q|\over2M}(-M\overline{\epsilon}_+\Gamma^a\gamma_0{}^i\epsilon_+),\qquad
{Q^a\over2M}(M\overline{\epsilon}_+\half\epsilon^{bcd}\hat
n^b\Gamma^{cd}\gamma_0{}^i\epsilon_+)
]\ ,\nonumber\\
\mu^{ij} &=& [
{Q^a\over2M}(-M\overline{\epsilon}_+\gamma_0{}^{ij}\epsilon_+),\qquad\quad
\epsilon^{abc}{Q^b\over2M}(-M\overline{\epsilon}_+\Gamma^c\gamma_0{}^{ij}
\epsilon_+)
]\ .
\label{eq:8dipole}
\end{eqnarray}
In particular, note the unavoidable presence of dipole moments
that are not parallel to the charge vector.

For the Kaluza-Klein gauge
fields (the left-hand entries in (\ref{eq:8dipole})), we find that the
magnetic dipole moment corresponds to $g=1$ as expected.  In addition to
the magnetic dipole moment, which is in the same Kaluza-Klein direction as
the charge, there is an electric dipole moment which is only non-vanishing
in the two directions orthogonal to the charge (since
$\hat n^ad_i^a[A_\mu]=0$).  This is indeed similar to the case of
four-dimensional $N=4$ BPS black holes, where graviphoton electric dipole
moments were found for electric black holes \cite{Duff:1997bs} (also with
the electric dipole moments orthogonal to the direction of the charge).

Turning to the three gauge fields $B_{\mu\,i}$, we see the interesting
result that a magnetic dipole moment is generated for the superpartners that
is not in the direction of the charge.  Since $\hat n^a\mu_{ij}^a[B_\mu]=0$,
only two of the three possible magnetic moments are non-zero.  Furthermore,
since the zero-mode structure of $\mu_{ij}^a[B_\mu]$ indicates that it is no
longer parallel to the spin $J_{ij}$, transition moments are necessarily
present.  Putting the
electric charge in the first Kaluza-Klein field, the various dipole moments
may be represented schematically as
\begin{eqnarray}
{\cal Q} &=& \hphantom{{|Q|\over2M}}[\matrix{Q\!\!&0&0,&0&0&0}]\ ,\nonumber\\
|d/{\cal S}| &=& {|Q|\over2M}[\matrix{0&1&1,&1&0&0}]\ ,\nonumber\\
|\mu/J| &=& {|Q|\over2M}[\matrix{1&0&0,&0&1&1}]\ .
\end{eqnarray}
This shows that the full 0-brane supermultiplet, in addition to having
$g=1$, in fact has dipole moment couplings to all six graviphotons,
regardless of which charge is originally turned on.  The dipole moments are
split to give three electric and three magnetic moments, which is suggestive
of the $SL(2)$ part of the $U$-duality group having some identification with
electric/magnetic duality.

Although we have focused on a Kaluza-Klein 0-brane in the
eight-dimensional theory, the above results must hold for all 0-brane
solitons as long as $U$-duality is valid.  We have indeed verified this
for the case of a 0-brane charged under $B_{\mu\,i}$.  For such a solution
we need to consider the ``dual'' projection $\widetilde{P}_\pm={1\over2}
(1\pm\gamma^{\overline{0}}\epsilon^{abc}\hat n^a\Gamma^{bc})$, resulting in
a 0-brane with similar properties to the above, with the main difference
being $\phi\to-\phi$ in this case.

Finally, the reader might be wondering what role is played by conventional
{\it rotating} black hole solutions of the Kerr type, carrying bosonic angular
momentum. This was discussed in \cite{Duff:1997bs,Duffrahmfeld2} in
the context of identifying four-dimensional string states and extreme black
holes.  The answer is that the Kerr angular momentum provides the {\it
superspin} $L$ of the supermultiplet in question. Thus $N=8$ KK states
belong to an $L=0$ supermultiplet and hence the spins of all members of
the multiplet come purely from fermionic hair. This is consistent with their
interpretation as the Dirichlet $0$-branes of the present paper, in which we
perform a further $T^{6}¥$ compactification from $D=10$ to $D=4$. $N=4$ KK
states, on the other hand, belong to an $L=1$ supermultiplet
and hence the spins of the members of the multiplet are a combination
of Kerr angular momentum and fermionic hair. Similar remarks apply
to $N=2$, for which $L=3/2$. For $N=0$, all the KK modes
derive their spin purely from Kerr angular momentum.

The gyromagnetic ratios of conventionally rotating black holes in
Kaluza-Klein theory were calculated in \cite{Gibbons:1986ac}, with the
result $g=2-v^2$ where $v$ is the boost velocity in the $x^5$ (internal)
direction.  Since KK states saturating the BPS bound correspond to $v=1$,
it follows that Kerr angular momentum also contributes $g=1$ to the
gyromagnetic ratio.  Thus the $g=1$ result for KK states is a universal one,
irrespective of the extreme black hole interpretation of the particle's spin.

\section{Conclusions}

In this paper we have calculated the gyromagnetic ratios of the
Dirichlet $0$-branes, regarding them as extreme Type $IIA$ black holes
whose spin is generated by the fermionic zero modes. We again find
$g=1$ consistent with their interpretation as Kaluza-Klein modes.  We
also note that, following the techniques discussed in
\cite{Duffrahmfeld2},
the Matrix model \cite{Banks} parton interpretation, in which a charge
$n$ $0$-brane is a zero-energy bound state of $n$ singly charged
$0$-branes, is also (trivially) consistent with $g=1$ since the tensor
product of two short representations each with $g=1$ contains another short
representation with $g=1$. To complete the picture it
would be interesting to verify $g=1$ also from the viewpoint of
surfaces of dimension $0$ on which open strings can end \cite{Polchinski}.

\bigskip
\noindent {\bf Acknowledgment.}
JTL wishes to thank S. Ferrara and W. Sabra for useful discussions.

\newpage

%\bibliographystyle{preprint}
%\bibliography{d0gyro}
%\end{document}

%%%%%%%%%%%%% include bibtex generated bibliography %%%%%%%%%%%%%%%

%%%%%%%%%%%%%%%%%% end bibliography %%%%%%%%%%%%%%%%%%%%%%%%%%%%%%%

\end{document}